\documentclass[reprint,twocolumn,showpacs,preprintnumbers,superscriptaddress,prb]{revtex4-1}

\usepackage{graphicx,bm,times,url}
\usepackage{amsmath,amsthm,amsfonts}
\usepackage{pifont}

\newcommand{\BoldVec}[1]{\mathchoice%
  {\mbox{\boldmath $\displaystyle     #1$}}%
  {\mbox{\boldmath $\textstyle        #1$}}%
  {\mbox{\boldmath $\scriptstyle      #1$}}%
  {\mbox{\boldmath $\scriptscriptstyle#1$}}%
}
\newcommand{\EQ}{\begin{equation}}
\newcommand{\EN}{\end{equation}}
\newcommand{\EQA}{\begin{eqnarray}}
\newcommand{\ENA}{\end{eqnarray}}

\newcommand{\Fig}[1]{Fig.~\ref{#1}}

\newcommand{\xx}{\BoldVec{x}{}}
\newcommand{\yy}{\BoldVec{y} {}}

\newcommand{\BB}{\BoldVec{B} {}}

\newcommand{\AAA}{\BoldVec{A} {}}

\newcommand{\kk}{\BoldVec{k} {}}

\newcommand{\nab}{\BoldVec{\nabla} {}}

\newcommand{\grad}{{\rm grad} \, {}}

\newcommand{\dd}{{\rm d} {}}

\def\chir{\chi^{(2)}}
\def\chis{\chi^{[2]}}
\graphicspath{{./fig/}{./png/}}

\def\R{{\Bbb R}}

\def\1{{\bf 1}}

\def\grad{{\bf{grad}}}

\def\vol{{\rm vol}}

\def\R{{\Bbb R}}

\begin{document}

\title{Calculations for the Practical Applications of Quadratic Helicity in MHD}

\author{Petr M.\ Akhmet'ev}
\affiliation{National Research University Higher School of
Economics, Moscow, Russia}
\affiliation{IZMIRAN, Troitsk, Moscow region, Russia}

\author{Simon Candelaresi\footnote{Corresponding author.}}
\affiliation{Division of Mathematics, University of Dundee, Dundee, DD1 4HN, UK}

\author{Alexandr Yu Smirnov}
\affiliation{IZMIRAN, Troitsk, Moscow region, Russia}
\affiliation{National University of Science and Technology MISiS,
Moscow, 119049, Russia}


\begin{abstract}
For the quadratic helicity $\chir$ we present a generalization of the Arnol'd
inequality which relates the magnetic energy to the quadratic helicity,
which poses a lower bound.
We then introduce the quadratic helicity density using the
classical magnetic helicity density and its derivatives along magnetic field lines.
For practical purposes we also compute the flow of the quadratic helicity and show
that for an $\alpha^2$-dynamo setting it coincides with the flow of the
square of the classical helicity.
We then show how the quadratic helicity can be extended to obtain
an invariant even under compressible deformations.
Finally, we conclude with the numerical computation of $\chir$
which show cases the practical usage of this higher order topological invariant.
\end{abstract}


\maketitle

\section{Introduction}
Modern models of non-linear dynamos
(e.g.\ \cite{Frisch-Pouquet-Leorat-1975-JFluidMech, Ji1999PhRvL, VishniacCho2001, BrandenbDoblerSubramanian2002AN,
ssHelLoss09, GrahamBlackmanMinuteHelicity12, DelSordoTurbulence12, Telloni-Perri-2013-776-1-ApJ,
Pipin-2013-768-1-ApJ, Tziotziou-Moraitis-2014-570-L1-AA, Brandenburg-Petrie-2017-836-21-ApJ}),
laboratory plasmas (e.g.\ \cite{Taylor-1974-PrlE, Motojima2006LHD}),
the solar magnetic field
(e.g.\ \cite{Berger1984, Berger2000, KleeorinMoss2000, Demoulin-Berger-2003-215-203-SolPhys, Chae-2007-39-11-AdSpR,
Guo-Ding-2013-779-2-ApJ, Liu-Hoeksema-2014-785-13-ApJ, Romano-2014-794-118-ApJ, Pariat-Valori-2015-580-A128-AA})
and other magnetohydrodynamics (MHD) problems
(e.g.\ \cite{Kumar-Rust-1996-101-A7-JGRSP, ShukurovSokoloffSubramanian2006AA, helFluxGuerrero2010MNRAS,
Pevtsov-Berger-2014-186-1-SpSciRev, Russell-Yeates-2015-22-3-PhysPlasm, Linkmann-Dallas-2016-94-5-PRE})
are based on magnetic helicity conservation.
Magnetic helicity has a simple geometrical interpretation: by the Arnol'd theorem it measures
the pairwise asymptotic linking of magnetic field lines \cite{A-Kh}, i.e.\
it is an invariant of ideal MHD.

Such topological invariants are by their definition conserved under an ideal evolution
with vanishing magnetic resistivity where the magnetic field evolves under a Lie-transport
(e.g.\ \cite{Craig-Sneyd-1986-311-451-ApJ, mimetic14}).
Approximately ideal conditions exist in various astrophysical settings, like the solar corona
where the dynamics of the plasma is dominated by the magnetic pressure with negligible
contributions from the hydrostatic pressure gradients.

Additionally, in ideal MHD there exist several topological invariants, which are
deduced from the pairwise asymptotic linking numbers of magnetic field lines:
the quadratic helicities \cite{A} and higher momenta of helicity
\cite{KomendarczykThirdOrderHelicityA2009}.
It has been suggested
that such topological invariants give rise to additional constraints on the
evolution of a magnetized plasma (e.g.\ \cite{fluxRings10, Yeates_Topology_2010, knotsDecay11}).

We investigate the properties of these invariants and, in analogy to the
magnetic helicity, try to answer the following question:
``Is it possible to use the quadratic helicities as non-linear restrictions
in MHD problems?''.
Our results can be used to to further study e.g.\ the application of
higher momenta of helicity in describing nonlinear
dynamo saturation, mentioned in \cite{S-I-A}.
Here we give the definitions and properties of the quadratic helicities and conclude
with a practical example for which we compute one of the helicities numerically.

Helicity density is a function on the 4-dimensional space of the ordered pairs of magnetic field lines
$\{(L_i, L_j)\}$, which is determined by the Arnol'd asymptotic linking number $h(L_i, L_j)$ \cite{ArnoldHopf1974}.
The quadratic helicities $\chi^{(2)}$, $\chi^{[2]}$ are the two (non-central) second momenta of
the function $h(L_i, L_j)$.
The definitions are given in section \ref{sec: helicities}.

In this paper we study only the quadratic helicity $\chi^{(2)}$, but in section \ref{sec: helicities},
we will refer to $\chis$ and results from \cite{A, S-I-A, S-A-I2}.
In section \ref{sec: Arnold Inequality} a generalized Arnol'd inequality for
the quadratic helicity is introduced.
There we prove that the upper bound of the quadratic helicity is well defined using only the magnetic field $\BB$.
In section \ref{sec: chir2} we propose a formula for $\chi^{(2)}$ using local data (the magnetic
helicity density $(\AAA, \BB)$ and its directional derivative along the vector $\BB$).
This formula is rather complicated, but its analogy to the formula for magnetic helicity provides
an excellent base for applying $\chi^{(2)}$ in practice.
Since $\chir$ is not conserved under a general smooth deformation (diffeomorphism)
in section \ref{sec: compressibility} we give a generalization of $\chir$ that is
invariant also under diffeomorphisms with homogeneous density change.
For the sake of practical applications, in section \ref{sec: numerics} we present the
numerical evaluation of $\chir$ for a linked magnetic field.
In section \ref{sec: helicity flow} we prove that the analog of the helicity
flux exists for the quadratic helicity.
This result shows that the quadratic helicity can be practically applied.

\section{Quadratic magnetic helicities for thin magnetic tubes}
\label{sec: helicities}
The quadratic helicities $\chi^{(2)}$, $\chi^{[2]}$ are defined for a magnetic field
$\BB$ inside a bounded domain $\Omega \subset \R^3$, where $\BB$ is tangent to the
boundary of the domain \cite{A, S-I-A}.
Assume that the magnetic field is represented by a large, but
finite number of magnetic tubes $\Omega_i$, then the quadratic helicity
$\chi^{(2)}$ is well-defined by the formula:
\begin{eqnarray}\label{110}
\chi^{(2)}(\BB)=\sum_{i,j,k} \frac{\Phi^2_i \Phi_j \Phi_k n(L_j,L_i)n(L_i,L_k)}{\vol(\Omega_i)},
\end{eqnarray}
where $L_i, L_j, L_k$ is a non-ordered collection of central lines of the (thin) magnetic tubes
$\Omega_i$, $\Omega_j$, $\Omega_k$ with magnetic fluxes $\Phi_i, \Phi_j, \Phi_k$
through their cross-sections.
In each collection of 3 tubes one magnetic tube $\Omega_i$ is singled out (marked).
Two collections of 3 tubes are different even if the tubes coincide,
but a different tube is singled out.
In equation \eqref{110} $n(L_j,L_i)$ and $n(L_i, L_k)$ are pairwise linking
coefficients of central lines of the corresponding magnetic tubes and $\vol(\Omega_i)$
are the volumes filled by the magnetic tubes.

With the same formalism, the quadratic helicity
$\chi^{[2]}$ is defined by the following formula where the sum is over non-ordered pairs
of thin magnetic tubes:
\begin{eqnarray}\label{12}
\chi^{[2]}(\BB)=\sum_{i,j} \frac{\Phi^2_i \Phi^2_j n^2(L_i, L_j)}{\vol(\Omega_i)\vol(\Omega_j)}.
\end{eqnarray}

We now clarify equation \eqref{110} by using the asymptotic ergodic Hopf
invariant of magnetic lines (for the definition of the asymptotic Hopf invariant
of magnetic lines see e.g.\ \cite{A-Kh}).
We assume, for simplicity, that the thin magnetic tubes
$\Omega_i$, $\Omega_j$, $\Omega_k$ consist of closed magnetic field lines, each
of which is defined as a parallel shift of the central line of the corresponding tube.
Assume that the absolute value $\vert \BB \vert$ of the magnetic field is constant along
each magnetic line.
Then the value of the asymptotic Hopf invariant
$h(L_i, L_j)$ of a pair of magnetic lines $L_i \subset \Omega_i$ and $L_j \subset \Omega_j$
is calculated by the formula:
\EQ
h(L_i,L_j) = \vert \BB_i \vert \vert \BB_j \vert n(L_i,L_j) \vert L_i \vert^{-1} \vert L_j \vert^{-1},
\EN
where $\vert L_i \vert$ and $\vert L_j \vert$ are lengths of the corresponding magnetic lines
and we use the relation $\vert \BB_i \vert S_i = \Phi_i$ for the magnetic flux with the cross sectional
area $S_i$ of the magnetic tube $\Omega_i$ and $\vol(\Omega_i)=\vert L_i \vert S_i$.

The quadratic helicity $\chi^{(2)}(\Omega_i;\Omega_j \cup \Omega_k)$ over
the domain $\Omega=\Omega_i \cup \Omega_j \cup \Omega_k$ with a marked magnetic
tube $\Omega_i$ is calculated as the result of integrating the function
$h(L_i,L_j)h(L_i,L_k)$ over the domain $\Omega$.
With the assumption that the function $h(L_i,L_j)h(L_i,L_k)$ is
constant in $\Omega$; as the result of the integration, we get the value:
\begin{eqnarray} \label{eq: chi2round in Omega}
\chi^{(2)}(\Omega_i;\Omega_j \cup \Omega_k) & = & \vert \BB_i \vert^2 \vert \BB_j \vert
\vert \BB_k \vert n(L_i,L_j) \nonumber \\
& & \times n(L_i,L_k) \vert L_i \vert^{-2} \vert L_j \vert^{-1}
\vert L_k \vert^{-1} \nonumber \\
& & \times \vol(\Omega_i)\vol(\Omega_j)\vol(\Omega_k).
\end{eqnarray}

The remaining two equalities for marked tubes $\Omega_j$, $\Omega_k$ are analogous.
After taking the sum and a corresponding transformation, equation \eqref{eq: chi2round in Omega}
coincides with the corresponding term in equation \eqref{110}.
Hence, the quadratic helicity $\chi^{(2)}$ (also the quadratic helicity $\chi^{[2]}$)
is defined analogously to the asymptotic Hopf invariant; the only difference is the following.
Instead of a Gaussian linking number, which is a combinatorial invariant of the order $1$
(in the sense of V.A.\ Vassiliev), one uses a combinatorial invariant of the order $2$.
Definitions and properties of finite-type invariants (Vassiliev invariants) of links can be
found in e.g.\ \cite{P-S}.

Let us note, that the values of formula  \eqref{12} are not changed with
respect to a subdivision of magnetic tubes into a collection of thinner tubes
$\Omega'_{i,k} = \gamma_{i,k}\Omega_i$ and $\Omega'_{j,l} = \gamma_{j,l}\Omega_j$,
with scaling factors $\gamma_{i, k}$ with $\sum_k \gamma_{i, k} = 1$.
For example, if a magnetic tube $\Omega_i$ is divided into $N_i$ thinner parallel
magnetic tubes $\Omega'_{i, k}$ then we get:
\begin{eqnarray}
\sum_k \vol(\Omega'_{i, k}) & = & \vol(\Omega_i) \nonumber \\
\sum_k \Phi'_{i, k} & = & \Phi_i \nonumber \\
n(L'_{i, k},L'_{j, l}) & = & n(L_{i}, L_{j}). \nonumber
\end{eqnarray}
With such a subdivision, equation \eqref{12} can be rewritten for an $N_i$ in the following way:
\begin{eqnarray}\label{3}
 \chi^{[2]}(\BB) & = & \sum_{i,j}\sum_{k,l} \frac{\Phi'^2_{i, k} \Phi'^2_{j, l}
n^2(L'_{i, k}, L'_{j, l})}{\vol(\Omega'_{i, k}) \vol(\Omega'_{j, l})} \nonumber \\
 & = & \sum_{i, j} n^2(L_i, L_j) \sum_{k} \frac{\Phi'^2_{i, k}}{\vol(\Omega'_{i, k})}
 \sum_{l} \frac{\Phi'^2_{j, l}}{\vol(\Omega'_{j, l})} \nonumber \\
 & = & \sum_{i, j} n^2(L_i, L_j) \sum_{k} \frac{\gamma^2_{i, k} \Phi^2_{i}}{\gamma_{i, k} \vol(\Omega_{i})}
 \sum_{l} \frac{\gamma^2_{j, l} \Phi^2_{j}}{\gamma_{j, l} \vol(\Omega_{j})} \nonumber \\
 & = & \sum_{i,j} \frac{\Phi^2_i \Phi^2_j n^2(L_i,L_j)}{\vol(\Omega_i)\vol(\Omega_j)},
\end{eqnarray}
The right hand side of the equation does not depend on the number of subdivisions $N_i$.

\section{The Arnol'd inequality for $\chir$ and its generalizations}
\label{sec: Arnold Inequality}
The following inequality:
\begin{eqnarray}\label{01}
U_{(2)}(\BB) \ge C \vert \chi(\BB) \vert, \quad U_{(2)}(\BB) = \int (\BB,\BB)\ \dd\Omega,
\end{eqnarray}
where $(.,.)$ denotes the scalar product, $\chi$ the magnetic helicity and $C$ a
positive constant, is called the Arnol'd inequality \cite{A-Kh}.
This inequality relates the magnetic energy
(on the left hand side) to the magnetic helicity (on the right hand side).
The constant $C>0$ does not depend on the magnetic field
$\BB$, but on the geometrical properties of the domain $\Omega$,
which is assumed to be a compact domain supporting $\BB$.

Using the model for $\BB$ from \cite{A-K-S} we prove the inequality
\begin{eqnarray}\label{eq: Uk_chi2}
U_{(k)}(\BB) \geq C \chi^{[2]}(\BB)
\end{eqnarray}
is not valid for $k < 1$ with an arbitrary fixed $C > 0$, with
$$U_{(k)}(\BB) = \int |\BB|^k \ \dd\Omega, \quad k > 0.$$

Consider a magnetic field $\BB$ inside a thin closed magnetic
tube $\Omega \subset \R^3$ without the thinner concentric
magnetic tube $\Omega_{\varepsilon}$ of radius $\varepsilon > 0$.
The magnetic field $\BB$ is tangent to both boundary components of $\Omega_{\varepsilon}$.
The component of $\BB$ that is parallel to the central line (which is cut-out) is constant
(it is equal to $1$).
The meridional component of $\BB$, around the central line, is proportional to
$r^{\alpha}$, where $r$ is the distance from a point to the central line.
The parameter $\alpha <0$ determines the intensity of the meridional component of
$\BB$, but makes no contribution to the integral magnetic flow
trough the cross-section of the magnetic tube $\Omega_{\varepsilon}$, which is
perpendicular the the central line.

In the limit $\lim_{r \rightarrow \infty}$ of equation \eqref{eq: Uk_chi2} with
$\alpha \in (0,-\infty)$, the right and left hand side of equation
\eqref{eq: Uk_chi2} tend to $+\infty$.
In the case of $k < 1$ there exists a value $\alpha$, for which the right hand
side of the inequality is infinite, but the left hand side is finite.

The quadratic helicity is not continuous with respect to $C^1$--small deformations of $\BB$.
In a magnetic tube ergodic domains of magnetic field lines are not stable with respect to
small $C^1$--deformations \cite{Kudr, E-P-T};
the quadratic helicity is destroyed by such deformations.
When ergodic domains of magnetic lines are destroyed inside the common ergodic domain $\Omega'$,
magnetic helicity remains fixed, the quadratic helicity
$\chi^{(2)}(\BB)$ decreases down to its lower bound, which is equal to the square
of the magnetic helicity normalized by the volume of the domain: $\frac{\chi^2(\BB)}{\vol(\Omega')}$.

Here we present a generalization of the Arnol'd inequality $(\ref{01})$, using the idea from
\cite{K-V}.
The new inequality estimates magnetic energies for $k=6$ and $k = \frac{3}{2}$
using the quadratic helicity $\chir$.

Let $\BB$ be a magnetic field in a bounded domain $\Omega \subset \R^3$, which is tangent
to the boundary of the domain.
The following inequality is then satisfied:
\begin{eqnarray}\label{222}
\left(\frac{\pi}{16}\right)^{\frac{2}{3}}U_{(6)}^{\frac{1}{3}}(\BB)
U_{(\frac{3}{2})}^{\frac{4}{3}}(\BB)
\ge \sup_{\BB';\varepsilon} \chi^{(2)}(\BB') \ge \chi^{(2)}(\BB),
\end{eqnarray}
where $\varepsilon >0$ is an arbitrary infinitesimally small positive constant,
$\sup_{\BB';\varepsilon} \chi^{(2)}(\BB')$ is the upper boundary over arbitrary magnetic fields $\BB'$, which are
$\varepsilon$-closed in $C^1$-topology to the initial magnetic field $\BB$.
The right hand side of the inequality \eqref{222} is an invariant under a smooth volume-preserving
transformation of the domain $\Omega$.

\proof{
Observe that
\begin{equation*}
\chi^{(2)}(\BB) \le \int(\AAA, \BB)^2\ \dd\Omega \le \int \BB^2 \rho^2\ \dd\Omega,
\end{equation*}
where
\begin{equation*}
\vert \vert \AAA(\xx) \vert \vert \le \rho(\xx) = \frac{1}{4\pi} \int_{\Omega} \frac {\vert\vert \BB \vert\vert}
{\vert \vert \xx - \yy \vert\vert^2}\ \dd \yy.
\end{equation*}
Using arguments of \cite{A-Kh} (III, proof of theorem 5.3, the Hardy-Littlewood inequality),
by H\"older's inequality with $p=\frac{1}{3}$ and $q=\frac{2}{3}$ we obtain:
\begin{eqnarray*}
\int \BB^2 \rho^2\ \dd\Omega \le U_{(6)}^{\frac{1}{3}}(\BB)
\left(\int \rho^3\ \dd\Omega\right)^{\frac{2}{3}} \\
\le \left(\frac{\pi}{16}\right)^{\frac{2}{3}}U_{(6)}^{\frac{1}{3}}(\BB)
\left(U(\BB)_{(\frac{3}{2})}\right)^{\frac{4}{3}}. \qed
\end{eqnarray*}
}

\section{Quadratic Magnetic helicity Density}
\label{sec: chir2}
The magnetic helicity is computed through the magnetic helicity density $(\AAA, \BB)$.
The quadratic helicities, however, admit no densities which makes the invariants
hard to calculate.
In this section we introduce the analogue of magnetic helicity density
for the quadratic helicity $\chi^{(2)}$.

Denote the magnetic helicity density by $h=(\AAA,\BB)$ in $\Omega$.
A magnetic line $L_i$ is equipped with the magnetic flow parameter $\tau$, this parameter
determines the magnetic flow in $\Omega$ generated by the vector field $\BB(\xx)$.

Along each magnetic line $L_i$ define the decomposition
$(\AAA(\tau),\BB(\tau)) = \bar{h} + \delta f(\tau)$, where $\bar{h}$ is a mean value of $(\AAA(\tau),\BB(\tau))$
along the line $L_i$
(which is well-defined for almost arbitrary magnetic lines $L_i$) and
$\delta f(\tau)$ is a variation with zero mean value.
The definition of $\bar{h}$ is clear for closed magnetic lines, for open magnetic lines the Arnol'd
approach is based on the Birkhoff theorem.

The quadratic helicity $\chi^{(2)}(\BB)$ is defined in \cite{A} as the result of the
integration of the functions $\bar{h}^2$ over the domain $\Omega$, where the function
$\bar{h}^2(\xx)$ is constant on each magnetic line $L_i$.
With that definition we can rewrite the expression for $\chi^{(2)}$ using the magnetic vector
potential $\AAA$ and avoid computing mutual linking numbers like $n(L_i, L_j)$.

We now calculate the square of the mean value of $(\AAA,\BB)$ on a magnetic field line
$L_i$ starting at point $\xx$ with limit $T$, $0 \le \tau \le T$ as
$\left(m_{\xx,T}[(\AAA,\BB)]\right)^2$ for sufficiently large $T$.
In the limit $T \to +\infty$ we obtain the mean value.
This makes $m_{\xx,T}[(\AAA,\BB)]$ the mean value of $(\AAA,\BB)$ on $L_i$
over the parameter $\tau$.
Then we integrate the function $(m_{\xx,T=+\infty}[(\AAA,\BB)])^2$ in the domain $\Omega$.
As result we obtain $\chi^{(2)}(\BB)$.
Obviously, $(m_{\xx,T=+\infty}[(\AAA,\BB)])^2$ does not change if we choose a different starting
point $\xx \in L_i$, because the mean value over an infinite magnetic line does not depend on the
integral over a finite segment.
and we can take $(m_{\xx,T=+\infty}[(\AAA,\BB)])^2$ in the formula.

This method has a major defect.
The integration is in general highly sensitive on the starting point $\xx$ of the
magnetic line.
This is especially true for chaotic magnetic field lines.
In case we integrate along a random collection of curves, which are uniformly
distributed in $\Omega$,
the integral coincides not with $\chi^{(2)}(\BB)$, but with $\chi^2(\BB)/\vol(\Omega)$,
because the integral tends to its minimal possible value.
With this the minimal value of $\chi^{(2)}$ coincides with the lower bound \cite{A}.

Here we present an alternative way to calculate $\chi^{(2)}(\BB)$.
In order to simplify the proofs we assume
that each magnetic line $L_i$ is closed, but the method can be applied to the general case.
For simplicity we now omit the indices $i$.
Consider the formula for the mean square of the magnetic helicity along a magnetic line
starting at position $\xx$:
\begin{eqnarray}\label{1}
m_{\xx,T}[(\bar{h}(\xx) + \delta f(\xx))^2]=\bar{h}^2 + m_{\xx,T}[\delta f^2].
\end{eqnarray}
Here, the term $2 m_{\xx,T} [\bar{h} \delta f]$ is trivial along lines and is omitted, because
$\bar{h}$ is a constant and $\delta f(\xx)$, $\xx \in L$ is the term with zero mean value.
The integral over the domain $\Omega$ of the left hand side of formula
$(\ref{1})$ is easy to calculate:
\begin{eqnarray}\label{22}
\iiint m_{\xx,T}[(\bar{h} + \delta f)^2]\ \dd\Omega =\iiint (\AAA,\BB)^2\ \dd\Omega,
\end{eqnarray}
because $\bar{h}(\xx) + \delta f(\xx) = (\AAA,\BB)\vert_{\xx}$, $\xx \in L$, and
$$\iiint m_{\xx,T}[(\AAA,\BB)^2]\ \dd\Omega = \iiint (\AAA,\BB)^2\ \dd\Omega.$$

Our goal is to calculate the integral
$$
\iiint m_{\xx,T}[\delta f^2]\ \dd\Omega.
$$

To calculate the term
$m_{\xx,T}[(\delta f)^2]$ 
we present the double integration over a subdomain in the Cartesian product of the magnetic line starting at $\xx$.
Consider a mean integral over $\tau \in [0,T]$:
\begin{eqnarray}\label{30}
\frac{1}{T} \int_0^T (\bar{h} + \delta f(\tau)) \int_0^{\tau}
\frac{\dd(\delta f(\tau_1))}{\dd\tau_1}\ \dd\tau_1\ \dd\tau,
\end{eqnarray}
where $\tau_1$ and $\tau$ are the curve parameters of the magnetic flow  with
$0 \le \tau_1 \le \tau \le T$ on $L$ starting at $\xx$.
By taking the average over $\tau$ we assume that $T \to +\infty$.


Using the equation:
\begin{equation}\label{41}
m_{\xx}\left[{\rm p.v.} \int^0_{-\infty} \frac{\dd(\delta f(\tau_1))}{\dd\tau_1}\ \dd\tau_1\right] = 0,
\end{equation}
where the integral over $\tau_1$ is calculated by its principal value (p.v.), or, as the Ces\`aro mean.
From $(\ref{41})$ we then get:
\begin{eqnarray}\label{42}
&& m_{\xx,\tau}[\delta f^2] = \nonumber \\
&& m_{\xx}\left[ \frac{1}{T} \int_0^T \int_{-\infty}^{\tau}(\bar{h} + \delta f(\tau))
\frac{\dd \delta f(\tau_1)}{\dd\tau_1}\ \dd\tau_1\ \dd\tau\right].
\end{eqnarray}
If we put $\tau_1 = \tau + \tau_2$ in the equation we obtain:
\begin{eqnarray}\label{43}
&&  m_{\xx,\tau}[\delta f^2] =  \nonumber \\
&& m_{\xx}\left[ \frac{1}{T} \int_0^T \int_{-\infty}^{0}(\bar{h} + \delta f(\tau))
\frac{\dd \delta f(\tau_2+\tau)}{\dd \tau_2}\ \dd\tau_2\ \dd\tau \right],
\end{eqnarray}
or, if we change the order of integration:
\EQA
 && m_{\xx,\tau}[\delta f^2] = \nonumber \\
 && m_{\xx}\left[ {\rm p.v.} \int_{-\infty}^{0} \frac{1}{T}\int_0^T(\bar{h} + \delta f(\tau)) \frac{\dd \delta f(\tau_2+\tau)}{\dd\tau_2}\ \dd\tau\ \dd\tau_2\right]. \nonumber
\ENA
The first term in the integration vanishes, i.e.
$$
\int_{-\infty}^{0} \frac{1}{T}
\int_{0}^{T} \bar{h} \frac{\dd \delta f(\tau_2 + \tau)}{\dd \tau_2}\ \dd \tau\ \dd \tau_2 = 0.
$$

Let us now define the function $\phi$ as
\begin{equation}
\phi(\tau_2, \xx) = \frac{1}{T} \int_{0}^{T} \delta f(\tau) \frac{\dd \delta f(\tau_2 + \tau)}{\dd \tau_2}\ \dd \tau.
\end{equation}
By the ergodicity we may replace in the integral $-\infty$ by $+\infty$
which results into
\begin{eqnarray}\label{44}
m_{\xx,\tau}[\delta f^2] & = &
m_{\xx} \left[{\rm p.v.} \int_{-\infty}^{0} \phi(\tau_2,\xx)\ \dd \tau_2\right] \nonumber \\
 & = & -m_{\xx} \left[{\rm p.v.} \int_0^{+\infty} \phi(\tau_2,\xx)\ \dd \tau_2\right].
\end{eqnarray}

Equation \eqref{44} is clear: for an arbitrary $\xx \in L$
define the function $\int_0^{\tau}  \delta f(\tau_2)\ \dd\tau_2 =
\varphi(\tau,\xx)$ for $\tau \ge 0$ with the boundary condition $\varphi(0,\xx)=0$.
This function is a branch of the scalar potential of the vector $\AAA$ along the magnetic
line $L$ with starting point $\xx$.
The function $\varphi(\tau,\xx)$ is well-defined by the boundary condition $\varphi(0,\xx)=0$.
Take the field $C(\xx)$, such that $m[\varphi(\tau,\xx) + C(\xx)]=0$ on $L$.
Denote $\varphi(\tau,\xx)+C(\xx)$ by $\bar{\varphi}(\tau,\xx)$.
With that, and because $C(\xx) = -m[\varphi(\tau,\xx)]$, we obtain
$C(\xx) = \bar{\varphi}(\tau,\xx) \vert_{\tau=0}$.

The equation \eqref{44} can be replaced by:
\begin{eqnarray}\label{45}
 && m_{\xx,\tau_2}[\delta f^2] = -m_{\xx,\tau}\left[\frac{1}{T} \int_{0}^{T} \bar{\varphi}(\tau,\xx)
\frac{\dd \delta f(\tau_2)}{\dd \tau_2}\ \dd\tau \right]  \nonumber\\
 && {\rm p.v.} [\bar{\varphi}]=0.
\end{eqnarray}
Then the formula for the quadratic helicity is presented as:
\EQ\label{47}
\chi^{(2)} = \iiint (\AAA,\BB)^2\ \dd \Omega - \iiint m_{\xx,\tau}[\delta f^2]\ \dd \Omega,
\EN
with the second term given by equation \eqref{44} or \eqref{45}.

Finally, let us prove the boundary condition for $\phi(\tau_2,\xx)$ in $(\ref{44})$:
\begin{eqnarray}\label{77}
\iiint \phi(0,\xx)\ \dd\Omega = 0.
\end{eqnarray}
Recall, $$(\AAA(\tau),\BB(\tau))=\bar{h}+\delta f(\tau),
$$
$$
\frac{\dd (\AAA(\tau),\BB(\tau))}{\dd \tau} =
\frac{\dd \delta f(\tau)}{\dd \tau}.
$$
We get $\phi(0,\xx)=(\AAA,\BB)\frac{\dd(\AAA,\BB)}{\dd\tau} =
\frac{1}{2} \frac{\dd(\AAA,\BB)^2}{\dd\tau}$ and
\begin{eqnarray}\label{48}
\iiint \frac{\dd(\AAA,\BB)^2}{\dd\tau}\ \dd\Omega = 0,
\end{eqnarray}
because the magnetic flow preserves the integral
$\iint (\AAA,\BB)^2\ \dd\Omega$.

Analogously, we get
$$ \iiint \frac{\dd^2 \phi(0, \xx)}{\dd\tau^2}\ \dd\Omega = \iiint(\AAA(\tau),\BB(\tau))\frac{\dd^3(\AAA(\tau),\BB(\tau))}{\dd\tau^3}\ \dd\Omega = $$
$$
\iiint 3\frac{\dd^2(\AAA(\tau),\BB(\tau))}{\dd\tau^2}\frac{\dd(\AAA(\tau)\BB(\tau))}{\dd\tau}\ \dd\Omega + $$
$$
\iiint(\AAA(\tau),\BB(\tau))\frac{\dd^3(\AAA(\tau),\BB(\tau))}{\dd\tau^3}\ \dd\Omega = $$
$$\iiint \frac{\dd^2}{\dd\tau^2}\left[(\AAA(\tau),\BB(\tau))\frac{\dd(\AAA(\tau),\BB(\tau))}{\dd\tau}\right]\ \dd\Omega = 0.$$

By analogous arguments, for $k \ge 0$, we get:
\begin{eqnarray}\label{77b}
\iiint \frac{\dd^{2k}\phi(0,\xx)}{\dd\tau^{2k}}\ \dd\Omega = 0.
\end{eqnarray}

By construction,
\begin{eqnarray}
\frac{\dd (\AAA(\tau),\BB(\tau))}{\dd\tau} = (\grad(\AAA,\BB),\BB). \nonumber
\end{eqnarray}
Denote
$$
\iiint \phi(\tau,\xx)\ \dd\Omega = \Phi(\tau),\quad \tau \ge 0.
$$
Let us decompose this function into the Fourier integral, using $(\ref{77})$, $(\ref{77b})$:
\begin{eqnarray}\label{8}
\Phi(\tau) = {\rm p.v.} \int\limits_0^{+\infty} b(y) \sin(y\tau)\ \dd y.
\end{eqnarray}
In this formula we collect all possible Fourier coefficients $b(y)$ over the spectrum $y$ with elementary harmonics
$\sin(y\tau)$, each elementary harmonic satisfies equation $\eqref{77b}$
and for $\tau \to +\infty$ tends to its principal value
$${\rm p.v.}\lim_{\tau \to +\infty} \frac{b(y)}{y}[1-\cos(y\tau)]=\frac{b(y)}{y}.$$
The required second term in $(\ref{47})$ is calculated as the principal value of
$\Phi(\tau)$, $\tau \to + \infty$ over $\tau$, or as the Ces\`aro mean value
$\left\langle \Phi\right\rangle$ of $\Phi(\tau)$ by the following formula:
\begin{eqnarray}\label{9}
\left\langle \Phi\right\rangle = \left[\int_0^{+\infty} \frac{b(y)}{y}\ \dd y\right].
\end{eqnarray}

Let us assume that the spectral densities
$b(y)$ have a compact support which belongs to the segment
$y \in [\delta_0,\delta_1]$, $\delta_0 >0$.
This assumption gives simplifications of the problem for
$y \to +\infty$ (very long complicated magnetic lines), and for $y \to 0+$ (higher harmonics of
the magnetic spectrum), in the case of closed magnetic lines with a lower estimation of curvature by a positive constant the condition is satisfied.
Let us calculate values, using equation \eqref{8},
$$  \frac{\dd \Phi(\tau)}{\dd\tau}, \quad \frac{\dd^3 \Phi(\tau)}{\dd\tau^3}, \quad
\dots, \quad \frac{\dd^{2k-1} \Phi(\tau)}{\dd\tau^{2k-1}}, \quad \dots $$
for $\tau=0$, $k \ge 1$ by the formulas:
\begin{eqnarray}\label{10}
\left.\frac{\dd^{2k-1} \Phi(\tau)}{\dd \tau^{2k-1}} \right|_{\tau=0} & = &
\iiint (\AAA,\BB) \left. \frac{\dd^{2k}(\AAA,\BB)}{\dd \tau^{2k}} \right\vert_{\tau=0}\ \dd\Omega \nonumber \\
& = & (-1)^k \int_{\delta_0}^{\delta_1}y^{2k-1} b(y)\ \dd y. \label{11}
\end{eqnarray}
From equation \eqref{10} we obtain $\Phi(0)=0$.
The left hand sides of the formulas are calculated by data, while the right hand sides
are the total collection of momenta of the required function $\frac{b(y)}{y}$ in the integral $(\ref{9})$.
By this collection the integrals \eqref{44}, and \eqref{9} are calculated as follows.

For the parameters $y \ge \delta_0$, $a>0$, ($a \gg \delta_0^{-2}$) we have the limit
$\lim_{a \to +\infty}{(1-\exp({-ay^2}))} = 1$.
But we can also write this expression using its Taylor expansion:
$$ 1 \approx 1 - \exp{(-ay^2)} = \sum_{k=1}^{\infty} (-1)^{k+1}\frac{a^k y^{2k}}{k!}. $$
With that the integral $(\ref{9})$ is calculated as following:
\begin{eqnarray}\label{astast}
\left\langle \Phi\right\rangle & = & \int_{\delta_0}^{\delta_1} \frac{b(y)}{y}\ \dd y \\
 & = & \sum_{k=1}^{\infty} (-1)^{k+1} \frac{a^k}{k!}\int_{\delta_0}^{\delta_1} y^{2k-1}b(y)\ \dd y \\
 & = & -\sum_{k=1}^{\infty} \frac{a^k}{k!} \left.\frac{\dd^{2k-1} \Phi(\tau)}{\dd \tau^{2k-1}} \right|_{\tau=0}.
\end{eqnarray}
This formula (of $(1-\exp{(-a)})$--type, $a>0$) is taken in the case $a$ is sufficiently large.


Let us consider the simplest example.
Assume that the magnetic field contains the closed magnetic line $L$.
The cyclic covering over this magnetic line is the real line $\tilde{L}=(-\infty,+\infty)$ with the period $2\pi$.
Assume that $(\AAA, \BB) = \bar{h} + \sin(\tau)$, $\tau \in \tilde{L}$.
Then the first term of the integral \eqref{47} is
$\bar{h}^2 + \frac{1}{2}$, the second term is $\frac{1}{2}$ and the quadratic helicity over $L$ equals to $\bar{h}^2$.

Let us calculate the second term in this formula using the function $(\ref{8})$. We get:
$$ -\left\langle \Phi \right\rangle = m_{x,\tau}[\delta f^2]=-m_{\tau}\left[\int_0^{+\infty} \Phi(\tau_2)\ \dd \tau_2\right] = $$
$$ -{\rm p.v.}\int_0^{+\infty} m_{\tau}\left[ \int_0^{+\infty} \sin(\tau) \cos(\tau + \tau_2)\ \dd \tau\right] \dd\tau_2 = $$
$$-\frac{1}{2} {\rm p.v.} \int_0^{+\infty} \sin(\tau_2)\ \dd\tau_2= -\frac{1}{2} {\rm p.v.}\lim_{\tau_2 \to +\infty} [1-\cos(\tau_2)] = -\frac{1}{2}. $$

Obviously, using the equation $(\ref{astast})$, because
$$\left. \frac{\dd^{2k}\Phi}{\dd\tau^{2k}}\right|_{\tau=0} =
\frac{(-1)^k}{2\pi} \int_0^{2\pi}\sin^2(\tau_2)\ \dd \tau_2,$$
$$\frac{1}{2\pi}\int_{0}^{2\pi} \sin^2(\tau_2)\ \dd\tau_2 = \frac{1}{2},$$ we get:
$$\left\langle \Phi \right\rangle = \lim_{a \to +\infty} -\frac{1}{2} \sum_{k=1}^{\infty} \frac{(-1)^k a^k}{k!} = \frac{1}{2}.$$



\section{Compressibility effects on the quadratic helicity $\chir$}
\label{sec: compressibility}
For many physical applications the underlying fluid or gas is described as
being compressible, rather than incompressible.
Such systems include those in plasma physics and MHD with
applications in astrophysics and fusion science.
While in their current definitions $\chir$ and $\chis$ are invariant under
incompressible fluid deformations, they need to be modified to assure
their invariance under a general diffeomorphism that includes compression.
We show that by simply including the fluid density we obtain quantities
that are invariant under a diffeomorphisms that change the density uniformly in space.

In \cite{A} (equation (2)) the authors use the square of the average magnetic helicity
along a magnetic line $\Lambda^{(2)}(T; \xx)$.
It is defined as
\begin{equation} \label{eq: Lambda_r}
\Lambda^{(2)}(T; \xx) = \frac{1}{T^2}\left( \int\limits_{0}^{T}
(\dot{\xx}(\tau), \AAA(\xx(\tau)))\ \dd\tau \right)^2,
\end{equation}
with the magnetic vector potential $\AAA$ of the field and the velocity vector
$\dot{\xx}(\tau) = \BB(\xx(\tau))$.
The integral is taken along a magnetic field line starting at position $\xx = \xx(\tau = 0)$.
With that \cite{A} wrote the quadratic helicity $\chir$ as
\begin{equation}\label{eq: chir aaa}
\chir = \limsup\limits_{T\rightarrow\infty} \int\Lambda^{(2)}(T; \xx)\ \dd D,
\end{equation}
where $D \in \mathbb{R}^3$ is a ball of radius $r$.
By Birkhoff Theorem the limit in $(\ref{eq: Lambda_r})$ exists for almost arbitrary $\xx$,
see \cite{A}, section 5.
The velocity $\dot{\xx}(\tau)$ along the magnetic field line is equivalent to $\BB$
at position $\xx(\tau)$.

Assume that magnetic lines are closed, this gives a simplification of the proof.
With that we can identify equation \eqref{eq: Lambda_r} as the squared of the average
magnetic helicity density along a magnetic flux line of integration length $T$.
For the sake of a compact notation we will write the integrand of equation
\eqref{eq: Lambda_r} as the scalar product of the magnetic vector potential and
the magnetic field, i.e.\ $(\AAA, \BB) \Leftrightarrow (\dot{\xx}(\tau), \AAA(\xx(\tau)))$.

We now investigate the cases of diffeomorphisms stretching the coordinate
system along and across the magnetic field lines assuming a
fluid density of $\rho_0 = 1$ before applying the mapping.
For a stretching along the magnetic field lines by a factor of $\lambda$, the
density changes to $\rho = \lambda^{-1}$.
$\BB$ is invariant under such a transformation because the integral magnetic flow
is invariant and the cross-section is fixed.
The total length of the field line at parameter $\tau = T$ changes,
but the mean of the magnetic helicity density does not, hence
$\overline{(\AAA, \BB)} \mapsto \overline{(\AAA, \BB)\rho^{-1}}$.
So, the function $\overline{(\AAA, \BB)\rho^{-1}}$ is frozen in.
We now substitute $(\AAA, \BB)$ by $(\AAA, \BB)\rho^{-1}$ in equation \eqref{eq: Lambda_r}
which adds a factor of $\rho^{-1}$ in equation \eqref{eq: chir aaa}:
\begin{eqnarray}\label{eq: Lambda_rho}
\Lambda_{\rho}^{(2)}(T; \xx) = \frac{1}{T^2}\left( \int\limits_{0}^{T}
\frac{(\dot{\xx}(\tau), \AAA(\xx(\tau)))}{\rho(\xx(\tau)))}\ \dd\tau \right)^2.
\end{eqnarray}
However, with $\rho$ also the measure $\dd D$ changes to $\rho \dd D$.
We get the following formula for $\chi^{(2)}$ in the compressed fluid:
\begin{equation}\label{eq: chir_rho}
\chi^{(2)}_\rho = \iint \Lambda^2_{\rho} \rho\ \dd D
\end{equation}

For a stretching across the magnetic field lines by a factor of $\lambda$
along both directions the density changes as $\rho = \lambda^{-2}$.
In order to conserve magnetic flux across comoving surfaces the magnetic field changes
according to $\lambda^{-2}\BB = \rho\BB$.
Again, $\overline{(\AAA, \BB)\rho^{-1}}$ is invariant.
The integral measure changes according to $\lambda^{-2}$.

In both cases the additional factor of $\rho$ in the two integrands cancel if $\rho$ does
not depend on space.
In that case we obtain a quadratic magnetic helicity that is invariant under (homogeneously)
density changing diffeomorphisms.

From the changes of $\chir$ for general transformations, we can conclude that
$\chir$ is not a function of the magnetic helicity $h$ and does form part
of invariants suggested by \cite{E-P-T}.
If it was true, then for two fields with the same helicity, the quadratic helicity would be the same.
Here we construct a simple counter example.
Take a field $\BB_1$ with helicity $h_1$.
Construct a second field $\BB_2$ via a topology conserving transformation of $\BB_1$,
then $h_1 = h_2$.
If this transformation is not volume conserving, then, in general, the quadratic
helicities are different, i.e. $h_1 \ne h_2$.

\section{Numerical calculations of $\chir_\rho$}
\label{sec: numerics}
For practical applications, like MHD simulations, equations \eqref{eq: Lambda_rho}
and \eqref{eq: chir_rho} can be computed numerically.
In order to evaluate the integral in equation \eqref{eq: Lambda_rho} we need to
trace magnetic field lines (streamlines), while for the integral in equation
\eqref{eq: chir_rho} we integrate in space.

\subsection{Hopf Link}
As simple test case we apply our calculations on the Hopf link which consists of two
magnetic flux tubes with a finite width which are interlinked.
In our construction they both have the radius $1$ and a tube diameter (FWHM) of $0.28284$.
Their centers are located at $(0, 0, 0)$ and $(1, 0, 0)$ with surface normals of
$(0, 0, 1)$ and $(0, 1, 0)$.
The magnetic vector potential for the first ring is given as
\begin{eqnarray}\label{eq: hopf link 1}
\zeta^1 & = & e^{\left(-{\left(x^{2} + y^{2} + z^{2} - 2 \, \sqrt{x^{2} + y^{2}} + 1\right)} {\eta}\right)}/\eta \nonumber \\
A^1_x & = & \frac{\sqrt{\pi} \sqrt{{\eta}} x {\zeta^1} \text{erf}\left(\sqrt{{\eta}} z\right) e^{\left({\eta} z^{2}\right)}}{2 \, \sqrt{x^{2} + y^{2}}} \nonumber \\
A^1_y & = & \frac{\sqrt{\pi} \sqrt{{\eta}} y {\zeta^1} \text{erf}\left(\sqrt{{\eta}} z\right) e^{\left({\eta} z^{2}\right)}}{2 \, \sqrt{x^{2} + y^{2}}} \nonumber \\
A^1_z & = & \frac{1}{2} \, {\zeta^1},
\end{eqnarray}
while for the second it is given as
\begin{eqnarray}\label{eq: hopf link 2}
\zeta^2 & = & e^{\left(-{\left({\left(x - 1\right)}^{2} + y^{2} + z^{2} - 2 \, \sqrt{{\left(x - 1\right)}^{2} + z^{2}} + 1\right)} {\eta}\right)}/\eta \nonumber \\
A^2_x & = & \frac{\sqrt{\pi} \sqrt{{\eta}} {\left(x - 1\right)} {\zeta^2} \text{erf}\left(\sqrt{{\eta}} y\right) e^{\left({\eta} y^{2}\right)}}{2 \, \sqrt{{\left(x - 1\right)}^{2} + z^{2}}} \nonumber \\
A^2_y & = & \frac{1}{2} \, {\zeta^2} \nonumber \\
A^2_z & = & \frac{\sqrt{\pi} \sqrt{{\eta}} z {\zeta^2} \text{erf}\left(\sqrt{{\eta}} y\right) e^{\left({\eta} y^{2}\right)}}{2 \, \sqrt{{\left(x - 1\right)}^{2} + z^{2}}},
\end{eqnarray}
with the total vector potential
\EQ\label{eq: hopf link total}
\AAA_{\rm Hopf} = \AAA^1 + \AAA^2.
\EN

\begin{figure}[t!]\begin{center}
\includegraphics[width=1\columnwidth]{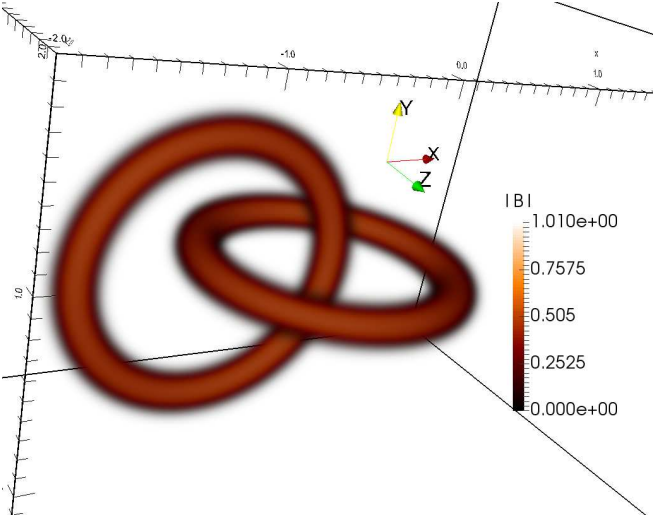}
\end{center}
\caption[]{
Volume rendering of the magnetic field strength for the Hopf link used in our calculations.
}
\label{fig: hopf link}
\end{figure}

From this we can write the magnetic field for the two flux rings as
\begin{eqnarray}
B^1_x & = & -\frac{\sqrt{\pi} {\eta}^{\frac{3}{2}} y z {\zeta^1} \text{erf}\left(\sqrt{{\eta}} z\right) e^{\left({\eta} z^{2}\right)} + {\eta} y {\zeta^1}}{\sqrt{x^{2} + y^{2}}} \nonumber \\
B^1_y & = & \frac{\sqrt{\pi} {\eta}^{\frac{3}{2}} x z {\zeta^1} \text{erf}\left(\sqrt{{\eta}} z\right) e^{\left({\eta} z^{2}\right)} + {\eta} x {\zeta^1}}{\sqrt{x^{2} + y^{2}}} \nonumber \\
B^1_z & = & 0
\end{eqnarray}
and
\begin{eqnarray}
B^2_x & = & \frac{\sqrt{\pi} {\eta}^{\frac{3}{2}} y z {\zeta^2} \text{erf}\left(\sqrt{{\eta}} y\right) e^{\left({\eta} y^{2}\right)} + {\eta} z {\zeta^2}}{\sqrt{x^{2} + z^{2} - 2 \, x + 1}} \nonumber \\
B^2_y & = & 0 \nonumber \\
B^2_z & = & -\frac{{\left(\sqrt{\pi} \sqrt{{\eta}} y \text{erf}\left(\sqrt{{\eta}} y\right) e^{\left({\eta} y^{2}\right)} + 1\right)} {\eta} {\left(x - 1\right)} {\zeta^2}}{\sqrt{x^{2} + z^{2} - 2 \, x + 1}}
\end{eqnarray}
with the total magnetic field as their sum
\EQ
\BB_{\rm Hopf} = \BB^1 + \BB^2.
\EN
A representation of this Hopf link is shown in \Fig{fig: hopf link}.

\subsection{Change of coordinates}
We apply five different simple homeomorphisms which change the grid density uniformly and compare
them to the initial grid.
By changing the grid we also need to transform the magnetic field $\BB$ and the magnetic
vector potential $\AAA$.
We do this by applying the pull-back on the corresponding differential forms:
\EQA
\alpha_0 & = & A_x\dd x + A_y\dd y + A_z\dd z \\
\beta_0 & = & B_x\dd y \wedge \dd z + B_y\dd z \wedge \dd x + B_z\dd x \wedge \dd y \\
\alpha & = & F^*(\alpha_0) \\
\beta & = & F^*(\beta_0),
\ENA
with the mapping $F$.

\begin{figure}[t!]\begin{center}
\includegraphics[width=1\columnwidth]{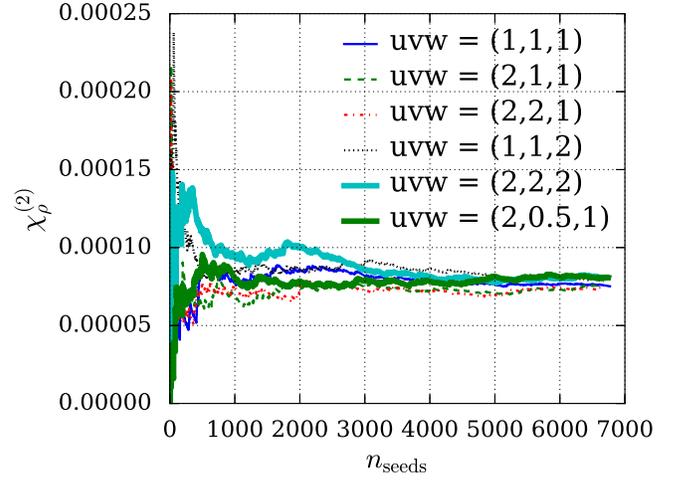}
\end{center}
\caption[]{
The modified quadratic helicity $\chir_\rho$ in dependence of the number of seed points for the
integration of equation \eqref{eq: chir_rho} for different homeomorphisms.
We clearly see the convergence to the same value of $\chir_\rho$.
Here we denote the mapping $F$ by the new coordinates $u$, $v$ and $w$, where e.g.\
uvw = (2, 0.5, 1) corresponds to $(u, v, w) = (2x, 0.5y, 1z)$.
}
\label{fig: chir rho}
\end{figure}

We stretch and compress the grid in various directions and perform the integral \eqref{eq: Lambda_rho}
and \eqref{eq: chir_rho} on the deformed field.
For simplicity we choose factors of $2$ for the deformations.
These deformations we the abbreviate using the short notation e.g.\ uvw $= (2, 0.5, 1)$,
which implies a transformation of $(u, v, w) = (2x, 0.5y, 1z)$, with the new
coordinates $(u, v, w)$.
In order to minimize selection bias we choose as starting points for our integration
a set of ca.\ $7000$ homogeneously distributed
points within a sphere of radius $2$ centered at the origin.
Those points are then transformed according to the homeomorphism.
A larger number of such points will provide us with a more accurate value for $\chir_\rho$.
In \Fig{fig: chir rho} we plot the value for $\chir_\rho$ for different grid deformations
and number of seed points for the integration.
We clearly see convergence to the same value which confirms our result that the modified
$\chir_\rho$ is invariant under a homeomorphism which homogeneously changes the grid density.

\section{Quadratic Magnetic Helicity Flow}
\label{sec: helicity flow}
By \cite{A}, Theorem 2, the quadratic magnetic helicity $\chi^{(2)}$ admits a continuous variation
in the case of $C^2$--small flows (vector of flows $\frac{\partial \BB}{\partial t}$ in
the domain $\Omega$ with its first and second partial derivatives small).
In \cite{E-P-T} is claimed, that  KAM-theory proves (for generic functionals) the analogous statement
for $C^k$--small flows, $k >3$.

Let us calculate, a contribution of the advection of magnetic lines
to the main term of the variation of $\chi^{(2)}$.
Let us consider $C^2$-small perturbations $\delta\BB$ and $\delta\AAA$:
$$\BB \mapsto \BB + \delta\BB, \quad \AAA \mapsto \AAA + \delta \AAA; \quad \nab\times \delta\AAA = \delta \BB.$$
Using Corollary 1 from \cite{A}, the first-order gauge transformation of the equation \eqref{eq: Lambda_r}
is represented by:
\begin{eqnarray*}
\Lambda^{(2)}(T;\xx) & \mapsto & \Lambda^{(2)}(T;\xx) + \\
 & & \frac{2}{T^2}\int_0^T (\dot{\xx}(\tau),\delta \AAA(\xx(\tau)))\ \dd\tau \\
 & & \times \int_0^T (\dot{\xx}(\tau),\AAA(\xx(\tau))\ \dd\tau + \\
 & & \frac{2}{T}(m_{\tau}[\delta \BB(\xx(\tau))]\vert_{\tau=T},\AAA(\xx(T))) \\
 & & \times \int_0^T (\dot{\xx}(\tau),\AAA(\xx(\tau))\ \dd\tau.
\end{eqnarray*}
The first extra term corresponds to the gauge term $\delta \AAA$.
The second extra term is given by the scalar product
of the average $m_{\tau}[{\delta \BB}]$ of the gauge vector $\delta \BB$
over the magnetic line $\xx(\tau)$ at the point $\xx(T)$.
(This term is given by the second term 
from the equation in Corollary 1 from \cite{A}.)
As the result, in the limit $\tau \to +\infty$ we get the following expression of the time-derivative of
the quadratic helicity:
\EQA
 && \frac{\dd \chi^{(2)}}{\dd t} = \nonumber \\
 && 2 \iiint m_{\tau} \left.\left[\left(\frac{\partial\AAA(\xx(\tau))}{\partial t}, \BB(\xx(\tau))\right)\right]
\right\vert_{\tau = \tau_0} + \nonumber \\
 && \left(m_{\tau} \left.\left[\frac{\partial \BB(\xx(\tau))}{\partial t}\right]\right\vert_{\tau=\tau_0},\AAA(\xx(\tau_0))\right) \nonumber \\
 && \times m_{\tau}[(\BB(\xx(\tau),\AAA(\tau))]\vert_{\tau = \tau_0}\ \dd\Omega. \nonumber
\ENA

By this property we get that the limit over the parameter
$T \to +\infty$ of formula \eqref{eq: Lambda_r} is uniform with respect to $t$--variations of the
magnetic field $\BB$.
To calculate $\frac{\dd \chi^{(2)}}{\dd t}$, for $t=t_0$, it is sufficient to calculate the
first-order derivative in equation \eqref{eq: Lambda_r} over $t$ for a prescribed $T$.
In particular, for $\alpha^2$-dynamos we get: $\frac{\partial \AAA}{\partial t}=\alpha \BB$;
and, assuming $\nab\times \BB = \kk\times\BB$, we have:
$\frac{\dd \chi^{(2)}}{\dd t} = 4\alpha k \chi^{(2)}.$
This means that the flow of quadratic magnetic helicity $\chi^{(2)}$
for the magnetic field with the $\kk$-vector
coincides with the flow of the square of the helicity $\chi^2$.

\section{Conclusions}
Both, the magnetic helicity and the quadratic helicities, can be calculated
from experimental data and be used in various MHD problems.
For that, the main tool and example is the Arnol'd inequality
\eqref{01}, which relates the geometry of magnetic field lines
(Gauss linking numbers or asymptotic Hopf invariants)
with the magnetic energy.
Analogous inequalities relate higher magnetic energies with
higher momenta of magnetic helicity (see the formula \eqref{222}).
This inequality shows that the upper bound of the quadratic helicity for
$C^0$-closed magnetic fields is limited.

For many problems in MHD one uses not only the total magnetic helicity, but its density.
The density of magnetic helicity is defined as $(\AAA,\BB)$.
This value is not invariant with respect to transformations of the domain with the magnetic field.
To calculate the quadratic helicity $\chi^{(2)}$ equation \eqref{47} is proposed.
The first term in the formula is easy to calculate.
To calculate the second term one needs to know the structure of the ergodic magnetic
subdomains in $\Omega$, where each subdomain contains a dense collection of magnetic field lines.
An ergodic subdomain can be shrunk to a surface, to a line, or even, to a critical point of $\BB$.
Therefore, quadratic magnetic helicities are much harder to use in applications then the magnetic helicity
and equation \eqref{eq: chir aaa} might be preferred.

We then showed that the quadratic helicity $\chir$ can be extended to be invariant under
non-volume preserving diffeomorphisms, as long as they change the density homogeneously.
We calculated that quantity numerically for the Hopf link and showed that it is indeed
invariant under such diffeomorphisms.
This has important applications in various fields of MHD.

\section*{Acknowledgements}

Petr Akhmet'ev and Alexandr Smirnov are supported in part by
the Russian Foundation
for Basic Research (grant No.\ 15-02-01407).
Simon Candelaresi acknowledges financial support from the UK's
STFC (grant number ST/K000993).
Petr Akhmet'ev is also supported by the Russian Foundation for Basic Research (Grant No.\ 17-52-53203 GFEN\_a).
Alexandr Smirnov is grateful to the financial support of the Ministry of Education and Science of
the Russian Federation in the framework of Increase Competitiveness Program of MISiS.
We thank the anonymous referee for the improvement suggestions.

\bibliography{references}
\end{document}